%% file: bad1243.tex
\documentclass[11pt]{article}

\usepackage{epsfig}
\usepackage{array}
\newcommand{\BABARPubYear}    {05}

\newcommand{\BABARConfNumber} {004}
\newcommand{\SLACPubNumber} {11313}
\newcommand{\LANLNumber} {0506065}

\input{babarsym}

\input{mysymbols}

\long\def\inst#1{\par\nobreak\kern 4pt\nobreak
    {\it #1}\par\vskip 10pt plus 3pt minus 3pt}

\begin{document}
{\pagestyle{empty}

\begin{flushright}
\babar-CONF-\BABARPubYear/\BABARConfNumber \\
SLAC-PUB-\SLACPubNumber \\
hep-ex/\LANLNumber \\
June 2005\\
\end{flushright}

\par\vskip 5cm

% Title of the paper
\begin{center}\Large \bf
Branching Fraction for $\Bp\to\piz\ellp\nu$, 
Measured in $\Y4S\to\BB$
Events Tagged by $\Bm\to\Dz\ellm\nub\parX$ Decays

\end{center}
\bigskip

\begin{center}
\large The \babar\ Collaboration\\
\mbox{ }\\
\today
\end{center}
\bigskip \bigskip

% Abstract
\begin{center}
\large \bf Abstract
\end{center}

We report a preliminary branching fraction of  
$(1.80\staterr{0.37}\systerr{0.23})\times10^{-4}$
for the charmless exclusive semileptonic $\Bp\to\piz\ellp\nu$ decay, 
where $\ell$ can be either a muon or an electron. This
result is based on data corresponding to an integrated luminosity of
81\invfb\ collected at the $\Upsilon(4S)$ resonance with the 
\babar\ detector. 
The analysis uses \BB\ events that are tagged by a
$B$ meson reconstructed in the semileptonic $\Bm\to\Dz\ellm\nub\parX$ decays,
where $X$ can be either a $\gamma$ or a $\pi^0$ from a $D^{*}$ decay.

\vfill
\begin{center}
Contributed to the 
XXII$^{\rm nd}$ International Symposium on Lepton and Photon Interactions at High~Energies, 6/30--7/5/2005, Uppsala, Sweden
\end{center}

\vspace{1.0cm}
\begin{center}
{\em Stanford Linear Accelerator Center, Stanford University, 
Stanford, CA 94309} \\ \vspace{0.1cm}\hrule\vspace{0.1cm}
Work supported in part by Department of Energy contract DE-AC03-76SF00515.
\end{center}

\newpage
} % end of pagestyle{empty}

\input{authors_lp2005}

\section{INTRODUCTION}

\begin{figure}[htbp]
\begin{center}
\epsfig{file=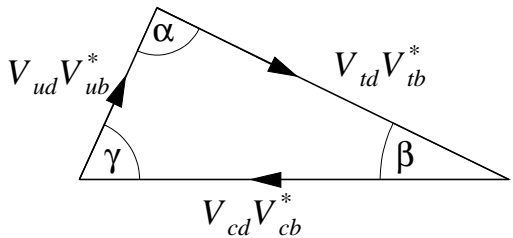,height=4cm}
\caption[Unitarity Triangle]{
\label{fig:triangle} Representation of the Unitarity Triangle}
\end{center}
\end{figure}

Measurements at $B$ Factories have significantly improved
our knowledge of \CP\ violation in the $\Bz-\Bzb$ system.
In particular, the angle $\beta$ of the Unitarity Triangle (see Figure~\ref{fig:triangle})
has been measured to a $5\%$ accuracy from time-dependent
\CP\ asymmetries in $b\to\ccbar s$ decays.

On the other hand,
experimental determination of the other two angles
and of the lengths of the two sides 
(with the third side normalized to unit length)
have yet to achieve comparable precision.
The uncertainty in the length of the side opposite 
to the angle $\beta$ is dominated by
the smallest CKM matrix element, $\Vub$.
Improved determination of $\Vub$ therefore translates directly
to a more stringent test of the Standard Model prediction.

Charmless semileptonic decays of $B$ mesons provide the
best probe for $\Vub$.
Measurements can be done either exclusively or inclusively, i.e., 
with or without specifying the hadronic final state.
Since both approaches suffer from significant theoretical uncertainties, it is
important to pursue both types of measurements and test their consistency.

The exclusive $B\to X_u\ell\nu$ decay rates
are related to $\Vub$ through form factors (FF).
In the simplest case of $B\to\pi\ell\nu$, the differential
decay rate (assuming massless leptons) is given by
\begin{equation}
  \frac{d\Gamma(\Bz\to\pim\ellp\nu)}{dq^2} =
  2\frac{d\Gamma(\Bp\to\piz\ellp\nu)}{dq^2} =
  \frac{G_F^2\Vub^2}{24\pi^3}|f_+(q^2)|^2p_{\pi}^3,
\end{equation}
where $G_F$ is the Fermi constant, 
$q^2$ is the invariant-mass squared of the
lepton-neutrino system and $p_\pi$ is the 
pion momentum in the $B$ frame. The FF $f_+(q^2)$ is 
calculated with a variety of approaches. Major improvements 
achieved recently in the calculation of the FF, 
based on light-cone sum rules~\cite{Ball05} and unquenched lattice 
QCD~\cite{HPQCD04,FNAL04} 
calculations, should allow a competitive determination of $\Vub$ 
using exclusive semileptonic
decays.

In this paper, we use the ISGW2 model~\cite{ISGW2}.
Alternate calculations~\cite{Ball05,HPQCD04,FNAL04} are considered to estimate 
model dependence of the result.
Measurements of the branching fraction
$\BR(B\to\pi\ell\nu)$ have been
reported by CLEO~\cite{CLEOpilnu}, Belle~\cite{Bellepilnu},
and \babar~\cite{nureco}.
The CLEO and \babar\ measurements use neutrino reconstruction in untagged \BB\ events; 
the Belle measurement uses semileptonic tags.
\babar\ has also reported a measurement of the total branching fraction
$\BR(B\to\pi\ell\nu)$ using fully-reconstructed hadronic tags~\cite{Breco}.

In this paper, we report a preliminary branching fraction 
measurement
from a study of the
$\Bp\to\piz\ellp\nu$ decay,
using event samples tagged by
$\Bm\to\Dz\ellm\nub\parX$ decays.\footnote{%
Charge-conjugate modes are implied throughout this paper.}
A similar study of the $\Bz\to\pim\ellp\nu$ decay is reported in a
separate paper~\cite{pilnu}.

We look for combinations of a $\Dz$ meson and a charged lepton
($e^-$ or $\mu^-$) that are kinematically consistent with
$\Bm\to\Dz\ellm\nub\parX$ decays.
For each such $B$ candidate, we define the signal side as
the tracks and neutral clusters that are not associated with
the candidate.
We search in the signal side for a signature of a $\Bp\to\piz\ellp\nu$ decay.
We take advantage of the simple kinematics of the $\Bp\to\piz\ellp\nu$
process and extract the
signal yield.
We calculate the branching fraction
using the signal efficiency predicted by a Monte Carlo (MC) simulation.
We correct for the data-MC efficiency differences using
control samples in which both $B$ mesons decay to tagging modes.

\section{THE \babar\ DETECTOR AND DATASET}

This measurement uses the $\epem$ colliding-beam data collected with the
\babar\ detector~\cite{BABAR} at the \pep2\ storage ring.
The data sample analyzed contains 88 million $\epem\to\BB$ events,
where \BB\ stands for $\Bp\Bm$ or $\Bz\Bzb$,
which corresponds to an integrated luminosity of 
81\invfb\ on the $\Upsilon(4S)$ resonance.
In addition, a smaller sample (10\invfb)
of off-resonance data recorded approximately $40\mev$ below the resonance 
is used for background subtraction and validation purposes.

We also use several samples of simulated $\BB$ events
to evaluate the signal and background efficiencies.
Charmless semileptonic decays $\BtoXulnu$ are simulated as a
mixture of exclusive channels
($X_u = \pi$, $\eta$, $\eta'$, $\rho$, and $\omega$)
based on the ISGW2 model~\cite{ISGW2} and
decays to non-resonant hadronic states~\cite{DFN}.

\section{ANALYSIS METHOD}\label{sec:analysis}

The event selection has been developed blind,  
that is, the analysis was first optimized using MC simulation
to obtain the largest expected statistical significance
of the signal yield; only then was it applied to data.

\subsection{Event Selection}

We search for candidate \BB\ events in which
one $B$ meson decayed as $\Bm\to\Dz\ellm\nub\parX$, 
$\Bzb\to\Dp\ellm\nub$ or $\Bzb\to\Dstarp\ellm\nub$, with 
$D^{*+}\rightarrow D^0 \pi^+$. 
The $D$ mesons are reconstructed in the
$\Dz\to\Km\pip$, $\Km\pip\pim\pip$, $\Km\pip\piz$,
and $\Dp\to\Km\pip\pip$ channels.
Charged $B$ tags are used to select the $\Bp\to\piz\ellp\nu$ decay mode.
The $D^0$ candidates from charged $B$ tags
are kept if their reconstructed mass lies within $10\sigma$ of the fitted mean of the $D^0$ 
mass distribution, where $\sigma$ is the experimental resolution. 
The signal region corresponds to $|m_{\Dz}^{\rm{reco}}-m_{\Dz}^{\rm{fitted}}|\le3\sigma$.
We define sideband regions,
which are used to subtract the combinatorial background,
as $3\sigma<|m_{\Dz}^{\rm{reco}}-m_{\Dz}^{\rm{fitted}}|\le10\sigma$.
The neutral $\Bz$ 
tags are used to reject $\Bz\Bzb$ events which are a 
background for the present analysis. In this case, a tighter $D^0$ mass 
selection, within $3\sigma$, is applied. The same mass window is used for 
$D^{\pm}$. The $\sigma$ values vary between $5\mevcc$ and 
$13\mevcc$ depending on the $D$ decay channel.

Candidate $D^{*+}\rightarrow D^0 \pi^+$ decays are required to have a mass difference, 
$m_{D^{*+}}-m_{D^{0}}$, of $\pm$ 5 $\mevcc$  
around its expected value (145.4 $\mevcc$). When a $D^{0}$ meson is identified 
as being part of a $\Bzb\to\Dstarp\ellm\nub\parX$ tag, the corresponding 
$\Bm\to\Dz\ellm\nub\parX$ tag is dropped.
Note that $\Dstar\to D\pi^0/\gamma$
are not explicitly reconstructed.

The $D^{(*)}$ candidates are combined with
electrons or muons to identify a $B\to\DorDstar\lnu\parX$ candidate.
The lepton must have a center-of-mass momentum\footnote{%
  Variables denoted with a star ($x^*$) are measured in the
  $\Upsilon(4S)$ rest frame; others are in the laboratory
  frame}
$\pstarl>1.0\gevc$.
The charge of the lepton must be the same as that of the
kaon from the $D$ candidate.

Referring to the tag-side $\DorDstar\ell$ combination as the ``$Y$'' system,
we calculate the cosine of $\BY$, the angle between $\vecpstar_B$
and $\vecpstar_Y$ as
\begin{equation}
  \cos\BY =
  \frac{2\Estar_B\Estar_Y-m^2_B-m^2_Y}
       {2\pstar_B\pstar_Y}.
  \label{eq:cosby}
\end{equation}
Equation~(\ref{eq:cosby}) assumes that a $B\to\DorDstar\ell\nu\parX$
decay has been correctly reconstructed, and the only undetected
particle in the final state is the neutrino.
If that is the case, $\cosBY$ should be between $-1$ and $+1$ within
experimental resolution.
If the tag has been incorrectly reconstructed,
Equation~(\ref{eq:cosby}) does not give a cosine of a physical angle,
and $\cosBY$ is distributed more broadly.
We require $-2.5<\cosBY<+1.1$.
The asymmetric range chosen for the $\cosBY$ cut reflects
the fact that a large fraction of the signal events contain
a $\Bm\to\Dstarz\ellm\nub$ decay in which the soft $\piz$ or $\gamma$
was not accounted for.
Such events tend to populate 
the $\cosBY$ distribution
between $-2.5$ and $-1.0$.

There is often more than one remaining $\DorDstar\ell\nu\parX$ candidate
in an event.
When this happens, we select the candidate with
the smallest value of $|\cosBY|$.
The average number of $\DorDstar\ell\nu\parX$ candidates per signal event
is 1.2 according to the MC simulation.
If at this point the best candidate represents a $\Bz$ decay
(i.e., $\Bzb\to\DorDstarp\ellm\nub$), the event is discarded.

From each event with a $\Bm\to\Dz\ellm\nub\parX$ candidate,
we remove the tracks and the neutral clusters that compose
$\Dz\ellm$.
We then search for a $\Bp\to\piz\ellp\nu$ candidate
in the remaining part (signal side) of the event, which 
must contain an identified lepton and a
pair of photons from $\piz\to\gamma\gamma$ that satisfy $115\mevcc<m_{\gamma\gamma}<150\mevcc$.
The two leptons in a candidate event must be
oppositely charged.

In order to suppress non-\BB\ backgrounds, which have a more jet-like 
topology than \BB\ events,
we require the ratio $R_2$ of the second and zeroth Fox-Wolfram
moments~\cite{FW}, computed using all charged tracks and unassociated
neutral clusters in the event, to be smaller than 0.5.

In addition,
we require the momenta of
the $\piz$ and lepton to satisfy $|\pstarpiz|+|\pstarl|\ge2.6\gevc$.
This cut removes more than 95\% of the \BB\ background
events while retaining nearly the entire phase space of the
$\Bp\to\piz\ellp\nu$ signal. In the limit of massless leptons and $\pi^0$s, 
this cut corresponds to the kinematic limit for the neutrino energy, $E^*_{\nu} \le 2.7~\gev $.

Analogously to $\cosBY$,
we can calculate the cosine of $\BX$,
the angle between $\vecpstar_B$ and $\vecpstar_{\piz\ell}$ as
\begin{equation}
  \cos\BX =
  \frac{2\Estar_B\Estar_{\piz\ell}-m^2_B-m^2_{\piz\ell}}
       {2\pstar_B\pstar_{\piz\ell}}.
  \label{eq:cosbx}
\end{equation}
This variable, again, should be between $-1$ and $+1$ for the
signal events, and distributed broadly for the background.
We require $-1.1<\cosBX<+1.0$.

If a signal event has been correctly reconstructed in the
$\Bm\to\Dz\ellm\nub$ and $\Bp\to\piz\ellp\nu$ decays,
no other particles should be present.
In reality, such events often contain extra neutral particles.
Some of them come from soft $\piz$s and/or photons
from decays of $\Dstarz$ or heavier charmed mesons.
We identify the photons that may have come from
$\Dstarz\to \Dz\piz$ and $\Dz\gamma$ decays by combining them with 
the $\Dz$ meson candidate; if the combination satisfies
$m_{\Dz\gamma(\gamma)}-m_{\Dz}<150\mevcc$ and $\cos\theta_{BY'}<1.1$,
where $Y'$ stands for the $\Dz\gamma(\gamma)\ell$ system, the photons are
considered as a part of the tag system and 
removed from the signal system.

At this point, we require that the event contains
no charged tracks 
besides the $\Bm\to\Dz\ellm\nub\parX$ and $\Bp\to\piz\ellp\nu$ candidates ($\Tleft=0$).
We further require the total residual neutral
energy of the event in the center-of-mass system to be
less than $300\mev$ ($\Eleft\le300\mev$).

To simplify the signal extraction
procedure, only events with a single $\Bp\to\piz\ellp\nu$ candidate
passing all the selections are kept. 
Only 1.6$\%$ of the MC signal events have more than one such candidate.

\begin{figure}[htbp]
\begin{center}
\epsfig{file=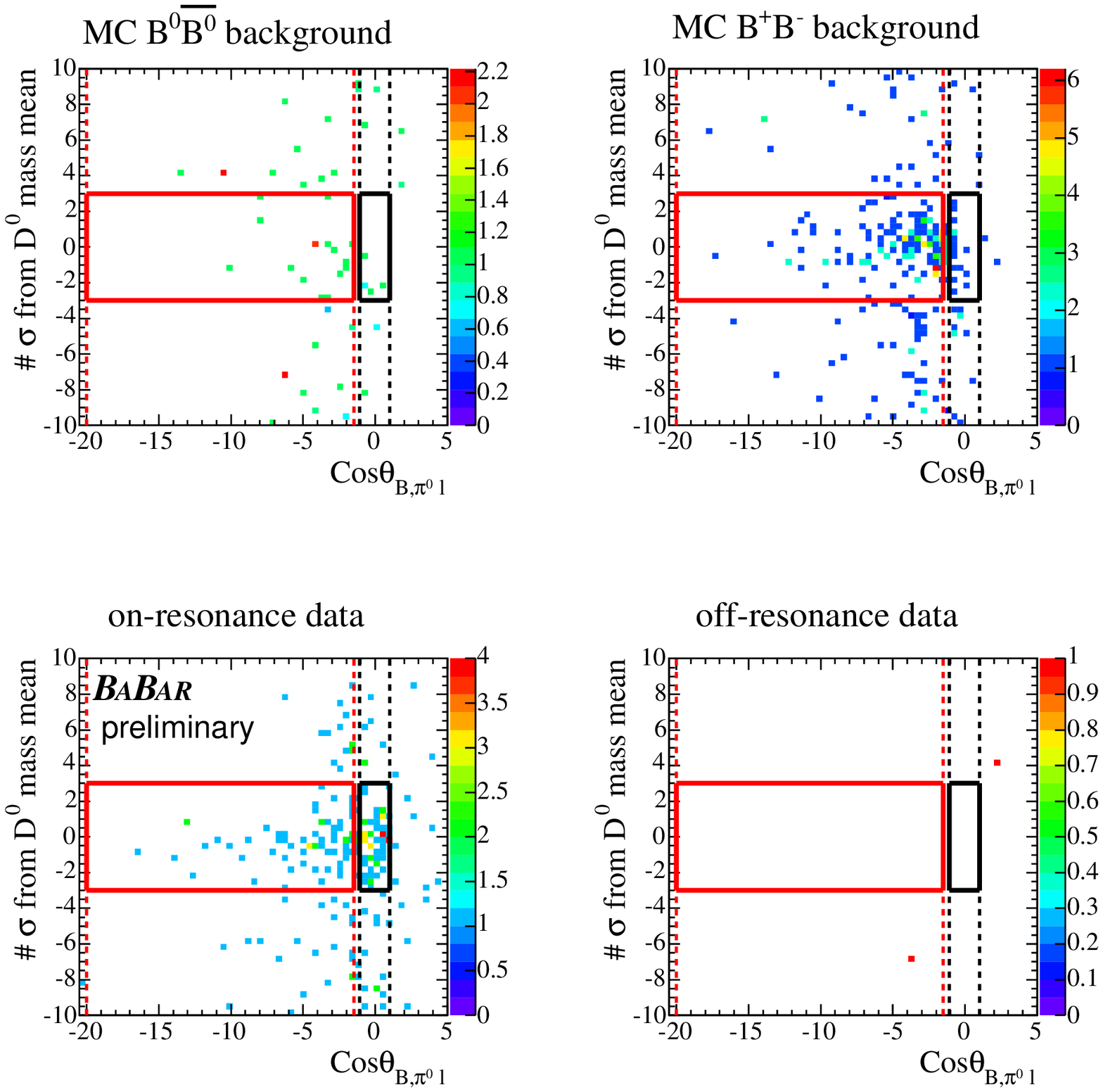, height=13cm}

\vspace{-0.2in}
\epsfig{file=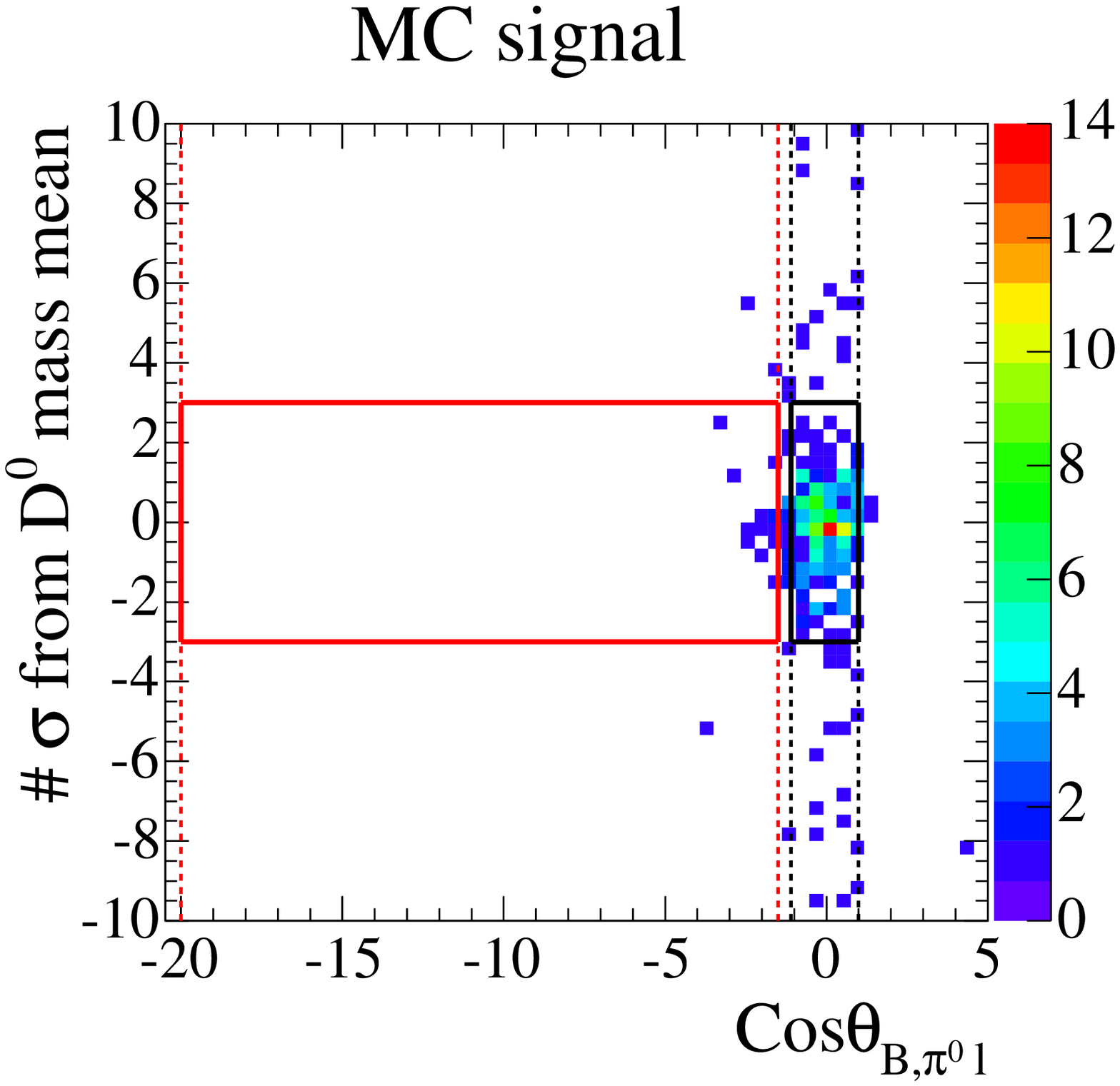, height=6.2cm}
\vspace{-0.3in}
\caption[Event distributions]{
\label{fig:RemainingEvents} Event distributions after all cuts (except on the variables shown),
 in the $\Dz$ mass -
$\cos\theta_{B,\pi^0 \ell}$ plane. The signal region is 
bounded by the black box: 
the reconstructed $\Dz$ mass values must be within 3$\sigma$ of their fitted mean ($\sigma$ is the experimental 
resolution on the $\Dz$ meson mass) and values of
$\cos\theta_{B,\pi^0 \ell}$ must lie between -1.1 and +1.0. The
$-20 \le \cos\theta_{B,\pi^0 \ell} \le -1.5$ region is defined as the $\cos\theta_{B,\pi^0 \ell}$
sideband while the $\Dz$-mass sideband regions correspond to 
$3\sigma<|m_{\Dz}^{\rm{reco}}-m_{\Dz}^{\rm{fitted}}|\le10\sigma$. 
The MC \BB\ and off-resonance 
distributions are not scaled to the integrated luminosity of the on-resonance data.}
\end{center}
\end{figure}

Distributions of the selected events in $\cos\theta_{B,\pi^0 \ell}$
and the difference between the reconstructed and fitted $\Dz$ mass 
are shown in Figure~\ref{fig:RemainingEvents}. These variables are used 
to extract the signal yield.

\subsection{Signal Yield}\label{ssec:yield}

We extract the total signal yield as
\begin{equation}
N = \Non - \Noff - \Nmc \frac{\Non^{SB}-\Noff^{SB}}{\Nmc^{SB}},
\end{equation}
where $\Non$, $\Noff$, and $\Nmc$ are
the numbers of events in the signal region
in the on-resonance data, off-resonance data, and simulated
\BB\ background, respectively, after the $\Dz$-mass sideband subtraction.
Note that this subtraction is performed with the assumption that the $D^0$ mass 
non-peaking background is linearly distributed over the range $\pm$10$\sigma$ 
around the nominal $D^0$ mass. The good fit of a straight line to the sideband 
distribution indicates that this assumption is reasonable.
The numbers $\Noff$ and $\Nmc$ are scaled according to the 
integrated luminosity of the on-resonance data.
The ``$SB$'' in superscript indicates a $\cosBX$ sideband region
defined as $-20\le\cosBX\le-1.5$. The ratio 
$(\Non^{SB}-\Noff^{SB})/\Nmc^{SB}$ is used to correct for the effect
of data/MC discrepancies in the $\Dz$-mass peaking background yield selected 
in the signal region $-1.1\le\cosBX\le1.0$. We assume that this ratio has 
the same value in the signal region since both regions are mostly populated by
events in which the tag-side decay is $\Bm\to\DorDstarz\ellm\nub$.
The ratio is found to be $\fexpression$ where the first error 
is statistical and the second error is a systematic uncertainty derived from the 
variation of the ratio when it is computed in three statistically independent bins of
the $-20 \le \cos\theta_{B,\pi^0 \ell} \le -1.5$ region.

We find no off-resonance events in the signal and $\Dz$ mass
sideband regions. The statistical uncertainty on these numbers, 
derived from the corresponding Poisson intervals, is large. Moreover, 
this uncertainty would have
to be scaled by a large factor ($\approx8$) to match the integrated 
luminosity of the on-resonance data.
This would result in a large error on the 
background subtraction, and calls for a special treatment.
Thus, we consider the alternative of computing the continuum
background yield using a factorized 
efficiency ($\epsilon_{\mathrm{MULT}}$)
equal to the product of the individual efficiencies of the last two cuts 
of the event selection, $\Tleft=0$ and 
$\Eleft\leq0.300 \gev$, which have a high power
of background suppression.
The direct use of $\epsilon_{\mathrm{MULT}}$ 
would be justified if these two cuts were uncorrelated. The variation between 
the usual efficiency and $\epsilon_{\mathrm{MULT}}$ has been studied as the
$\Tleft$ and $\Eleft$ cuts
were progressively relaxed. From the difference in the behavior of these two estimators,
a correction factor has been derived and a systematic uncertainty has been 
added to the multiplied efficiency: $\epsilon^{Corrected}_{\mathrm{MULT}} = (0.35 \pm 0.35)\epsilon_{\mathrm{MULT}}$. The resulting off-resonance event yield in the signal region, after the $\Dz$-mass sideband subtraction and scaled to the integrated luminosity of the on-resonance data, is $-0.3\pm2.5$.

\begin{figure}[htbp]
\begin{center}
\epsfig{file=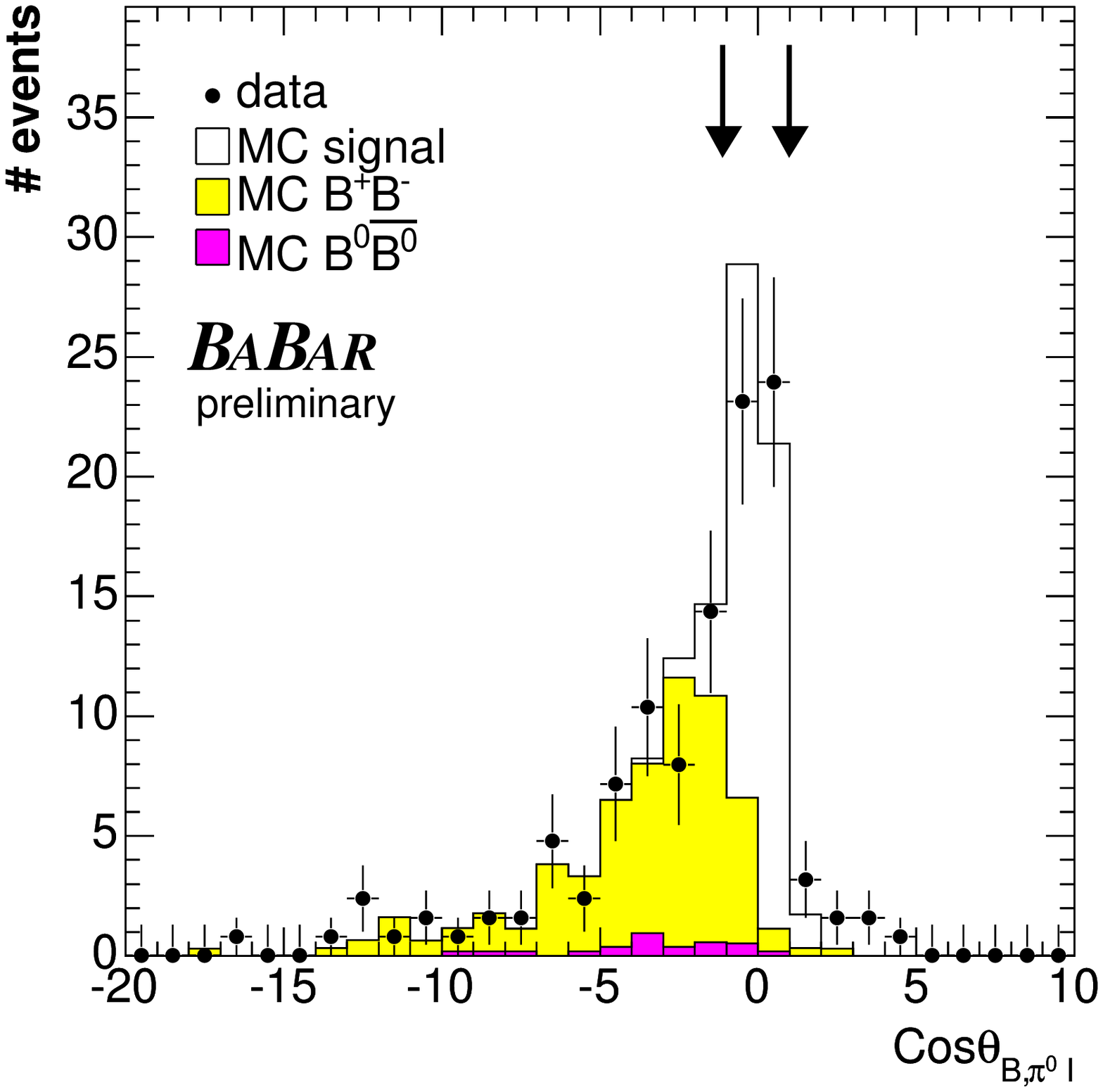,height=15cm}
\caption{\label{fig:pi0lnu}
$\cosBX$  distribution, after all other cuts, for events having the reconstructed 
$\Dz$ mass values within 3$\sigma$ of their fitted mean. The points are the on-resonance 
data and the error bars are statistical only. In order to illustrate the final 
yields and the significance
of the signal, each distribution (on-resonance data, MC signal, MC $\Bp\Bm$ and MC $\Bz\Bzb$)
is rescaled to the number of events found after 
the $D^0$-mass sideband subtraction. Note that the off-resonance data, 
due to their special treatment
(see Section \ref{ssec:yield}), are not shown.
The final $\cosBX$ selection is between -1.1 and 1.0, delimited by
the arrows.
}
\end{center}
\end{figure}

We find $\Non=52.0\pm8.1$ events in the signal region,
$-1.1<\cosBX<+1.0$, after the $\Dz$-mass sideband subtraction, of which
$6.7\pm2.0$, $0.6\pm0.9$, and $-0.3\pm2.5$ events are estimated
to be $\Bp\Bm$, $\Bz\Bzb$ and non-\BB\ backgrounds, respectively.
The errors are statistical. The signal yield is therefore $44.9\pm8.7$ events.
The corresponding yields before the $\Dz$-mass sideband subtraction
are $61\pm7.8$, $9.0\pm1.9$, $1.6\pm0.8$ and $1.5\pm1.7$ for on-resonance data,
$\Bp\Bm$, $\Bz\Bzb$ and non-\BB\ backgrounds, respectively.
Figure~\ref{fig:pi0lnu} illustrates the yields for both
data and MC samples and thus the significance of the
signal.

\subsection{Signal Efficiencies}\label{sec:eff}

The signal yield is expressed in terms of the efficiency $\effdata$, computed 
using a sum of $\Bp\to\piz e^+\nu$ and $\Bp\to\piz \mu^+\nu$ decays, as
\begin{equation}
  N = 2\effdata\BR(\Bp\to\piz\ellp\nu)N_B,
  \label{eq:effmat}
\end{equation}
where the factor of two comes from using electrons or muons in the branching fraction,
and $N_B$ is the number of $B^{\pm}$ mesons in the data sample.
Using the $\Upsilon(4S)\to\Bz\Bzb$ branching fraction
$f_{00}=\frac{1}{2}$, the number $N_B$
equals $2(1-f_{00})\NBB$,
where $\NBB$ is the number of \BB\ events in the data sample
and the factor of two comes from having two $B$ mesons in each event.

We use the signal Monte Carlo
simulation to provide an estimate of the total
efficiency:
\begin{equation}
\effMC = (1.40\pm0.10)\times10^{-3},
\end{equation}
The uncertainty is due to Monte Carlo statistics.

We evaluate the data-MC difference of the $\Bm\to\Dz\ellm\nub\parX$
selection efficiencies using the double-tag events,
in which both $B$ mesons decay to
$\Dz\ell\nu\parX$.
Properties of the $\Bm\to\Dz\ellm\nub\parX$ tags
such as the composition of the $\Dz$ decay channels
are similar between the tagged-signal and double-tag events.
The number of double-tag events is proportional to the
square of the tagging efficiency after subtracting the
small contribution from background.

The selection criteria for the double-tag events follow
the main analysis as closely as possible.
In each event, we look for two $\Dz\lnu\parX$ tags that do not overlap
with each other.
We remove all particles that are used in the tags and
require that there be no charged tracks.
The $\Eleft$ cut is increased
from 300\mev\ to 600\mev; this reflects the fact that the residual
neutral energy in a signal event comes from the tag-side $B$
decay, and a double-tag event contains two such decays.

From the study of the double-tag events
we extract the efficiency correction factor
\[
  \frac{\effdata}{\effMC} = 1.005\pm0.039.
\]
The error includes both the statistical and systematic uncertainties.
For the latter, we considered the difference in the results when
the selection criteria are varied, the residual background after
the $\Dz$-mass sideband subtraction, due
to possible non-linearities in the background vs. the reconstructed $Dz$ mass, and
the uncertainties in the exclusive $B\to X_c\lnu$ branching
fractions.

\section{SYSTEMATIC UNCERTAINTIES}\label{sec:systematics}

The significant sources of systematic uncertainties considered
and their impact on the measured branching fraction are
summarized in Table~\ref{tab:sys}.

\begin{table}[htbp]\centering
\caption{Fractional systematic errors on the branching fraction.}
\label{tab:sys}\smallskip
\begin{tabular}{l|R}
\hline\hline
Systematics & \sigma_{\BR}/\BR \\
\hline
MC \BB\ statistics                        & \pm4.6\% \\
MC signal statistics                      & \pm7.3\% \\
$(\Non^{SB}-\Noff^{SB})/\Nmc^{SB}$ ratio  & \pm6.0\% \\
Signal lepton tracking                    & \pm0.8\% \\
Signal lepton ID                          & \pm3.3\% \\
Signal $\piz$ efficiency                  & \pm5.0\% \\
Tagging efficiency                        & \pm3.9\% \\
$B\to\pi\ell\nu$ FF                       & \pm1.5\% \\
$\BR(B\to X_{c/u}\ell\nu)$ background     & \pm1.1\% \\
$\NBB$                                    & \pm1.1\% \\
\hline
Total                                     & \pm12.9\% \\
\hline
\end{tabular}
\end{table}

The systematic error is dominated by the
size of the available MC samples.
This includes both the signal MC sample for evaluating the
signal efficiency and the generic \BB\ MC samples for the
background subtraction.

The systematic uncertainties associated with the signal $\piz$ efficiency, 
the identification of the signal lepton and its tracking efficiency
are derived from detailed studies of the discrepancies in particle
reconstruction between MC simulation and data.
In particular, the systematic uncertainty in signal $\piz$ efficiency
is obtained by comparing the ratio of the $\tau^{-} \rightarrow h^{-} \piz \piz \nu$ yield to the $\tau^{-} \rightarrow h^{-} \piz \nu$ yield in data and simulation, using $e^+e^- \rightarrow \tau^+\tau^- $ events. 
The uncertainty associated with the tagging efficiency
was evaluated by using double-tag events (see Section~\ref{sec:eff}). 

For the $B\to\pi\lnu$ form factor, we used the ISGW2 model~\cite{ISGW2}. 
The related systematic uncertainty is evaluated
by reweighting~\cite{FFReweighting} the simulated signal events to reproduce 
the $q^2$ distribution predicted by the Ball-Zwicky, HPQCD and 
FNAL calculations~\cite{Ball05,HPQCD04,FNAL04}.
This uncertainty corresponds to the spread 
between calculations. The form factor 
affects the branching fraction through the
$q^2$ dependence of the signal efficiency.
Only the shape and not the normalization of the
form factor is relevant for the measurement of the branching
fraction.

Background events are mostly 
from $B \to X_c \ell \nu$ decays, with a small contribution from $B 
\to X_u \ell \nu$ decays. For many of these decays, the branching 
fractions are not well known. To evaluate
the associated uncertainty, the branching fractions of these decays were varied
simultaneously within their uncertainties~\cite{PDG04}. The resulting 
variation in the $\Bp\to\piz\ellp\nu$ branching fraction 
was taken to be the corresponding systematic uncertainty.

In addition to the studies discussed above, we performed crosschecks
to ensure that the simulation was adequately modeling the data close 
to the signal region,
before unblinding the data. In particular, we've studied
the data-MC agreement in sidebands of $\cosBX$ and $\Eleft$, and
observed no variation beyond expected statistical fluctuations.

\section{RESULTS AND SUMMARY}\label{sec:summary}

Using event samples tagged by $\Bm\to\Dz\ellm\nub\parX$ decays, we determine
the exclusive branching fraction $\BR(\Bp\to\piz\ellp\nu)$.
From the signal yield and the efficiencies evaluated in Section~\ref{sec:analysis},
we extract a preliminary result
\[
  \BR(\Bp\to\piz\ellp\nu) = (1.80\staterr{0.37}\systerr{0.23})\times10^{-4}.
\]

Table~\ref{tab:babarbf} summarizes the measurements of $\BR(B\to\pi\lnu)$
by the \babar\ collaboration.
Assuming isospin symmetry and the ratio of $B$ lifetimes
$\tau_{\Bp}/\tau_{\Bz}=1.081\pm0.015$~\cite{PDG04},
the measurements agree with each other
with $\chi^2=10.3$ for 5 degrees of freedom,
which corresponds to a one-sided probability of 7\%.  

We plan to extract $\Vub$ in the next 
version of this analysis, which will also include more data.

\begin{table}[htbp]\centering
  \caption{\label{tab:babarbf}
    Preliminary \babar\ measurements of $\BR(B\to\pi\lnu)$.
    The last row shows this measurement and the $\Bz\to\pim\ellp\nu$
    measurement in Ref.~\cite{pilnu}}
  \begin{tabular}{lCC}
  \hline\hline
  Technique & \BR(\Bz\to\pim\ellp\nu)\times10^4 & \BR(\Bp\to\piz\ellp\nu)\times10^4 \\
  \hline
  Neutrino reco.~\cite{nureco} & 1.41\pm0.17\pm0.20 & 0.70\pm0.10\pm0.10 \\
  Hadronic tag~\cite{Breco}    & 0.89\pm0.34\pm0.12 & 0.91\pm0.28\pm0.14 \\
  Semileptonic tag             & 1.03\pm0.25\pm0.13 & 1.80\pm0.37\pm0.23 \\
  \hline
  \end{tabular}
\end{table}

\section{ACKNOWLEDGMENTS}
\input{acknowledgements}

\end{document}

%% file: mysymbols.tex
\newcommand{\DorDstar}{D^{(*)}}
\newcommand{\DorDstarp}{D^{(*)+}}

\newcommand{\DorDstarz}{D^{(*)0}}
\newcommand{\parX}{(X)}
\newcommand{\pstar}{p^*}
\newcommand{\vecpstar}{\mathbf{p}^*}
\newcommand{\Estar}{E^*}
\newcommand{\pstarl}{\pstar_{\ell}}

\newcommand{\pstarpiz}{\pstar_{\pi^0}}
\newcommand{\BY}{\theta_{BY}}
\newcommand{\BX}{\theta_{B\piz\ell}}
\newcommand{\cosBY}{\cos\BY}
\newcommand{\cosBX}{\cos\BX}

\newcommand{\eff}{\varepsilon}
\newcommand{\effMC}{\eff^{\mathrm{MC}}}
\newcommand{\effdata}{\eff^{\mathrm{data}}}

\newcommand{\Non}{N_{\mathrm{on}}}
\newcommand{\Noff}{N_{\mathrm{off}}}
\newcommand{\Nmc}{N_{\mathrm{MC}}}

\newcommand{\NBB}{N_{\BB}}

\newcommand{\staterr}[1]{\pm{#1}_{\mathrm{stat.}}}
\newcommand{\systerr}[1]{\pm{#1}_{\mathrm{syst.}}}

\newcommand{\BtoXulnu}{B\to X_u\lnu}
\newcommand{\lnu}{\ell\nu}
\newcommand{\Eleft}{\mathrm{\it{E_{extra}}}}
\newcommand{\Tleft}{\mathrm{\it{T_{extra}}}}

\newcommand{\etal}{\textit{et al.}}
\newcommand{\PRL}[3]{Phys.\ Rev.\ Lett.\ \textbf{#1}, #2 (#3)}
\newcommand{\PRD}[3]{Phys.\ Rev.\ D \textbf{#1}, #2 (#3)}
\newcommand{\PLB}[3]{Phys.\ Lett.\ B \textbf{#1}, #2 (#3)}
\newcommand{\NIMA}[3]{Nucl.\ Instrum.\ Meth. A \textbf{#1}, #2 (#3)}
\newcommand{\JHEP}[3]{JHEP #1, #2 (#3)}
\newcommand{\fexpression}{1.02 \pm 0.11 \pm 0.35}

\newcolumntype{L}{>{$}l<{$}}
\newcolumntype{R}{>{$}r<{$}}
\newcolumntype{C}{>{$}c<{$}}

%% file: authors_lp2005.tex
\begin{center}
\small

The \babar\ Collaboration,
\bigskip

B.~Aubert,
R.~Barate,
D.~Boutigny,
F.~Couderc,
Y.~Karyotakis,
J.~P.~Lees,
V.~Poireau,
V.~Tisserand,
A.~Zghiche
\inst{Laboratoire de Physique des Particules, F-74941 Annecy-le-Vieux, France }
E.~Grauges
\inst{IFAE, Universitat Autonoma de Barcelona, E-08193 Bellaterra, Barcelona, Spain }
% ...

\end{center}\newpage

%% file: acknowledgements.tex
We are grateful for the 
extraordinary contributions of our \pep2\ colleagues in
achieving the excellent luminosity and machine conditions
that have made this work possible.
The success of this project also relies critically on the 
expertise and dedication of the computing organizations that 
support \babar.
The collaborating institutions wish to thank 
SLAC for its support and the kind hospitality extended to them. 
This work is supported by the
US Department of Energy
and National Science Foundation, the
Natural Sciences and Engineering Research Council (Canada),
Institute of High Energy Physics (China), the
Commissariat \`a l'Energie Atomique and
Institut National de Physique Nucl\'eaire et de Physique des Particules
(France), the
Bundesministerium f\"ur Bildung und Forschung and
Deutsche Forschungsgemeinschaft
(Germany), the
Istituto Nazionale di Fisica Nucleare (Italy),
the Foundation for Fundamental Research on Matter (The Netherlands),
the Research Council of Norway, the
Ministry of Science and Technology of the Russian Federation, and the
Particle Physics and Astronomy Research Council (United Kingdom). 
Individuals have received support from 
CONACyT (Mexico),
the A. P. Sloan Foundation, 
the Research Corporation,
and the Alexander von Humboldt Foundation.